\documentclass[superscriptaddress,onecolumn,floats,10pt,floatfix,nobibnotes,notitlepage,showkeys]{revtex4-1}

\usepackage{amsmath}
\usepackage{amsfonts}
\usepackage{amssymb}
\usepackage{graphicx}
\usepackage{color}
\usepackage[colorlinks=true,citecolor=blue,linkcolor=blue,urlcolor=black]{hyperref}
\usepackage{times}
\usepackage{amstext}
\usepackage{latexsym}
\usepackage{bbm}
\usepackage[normalem]{ulem}

\providecommand{\be}{\begin{equation}}
\providecommand{\ee}{\end{equation}}
\providecommand{\ba}{\begin{eqnarray}}
\providecommand{\ea}{\end{eqnarray}}

\usepackage{mathbbol}
\usepackage{mathtools}

\begin{document}
\title{Two-mode squeezing operator in circuit QED}

\author{E. C. Diniz}
\email{emanuelcardosodiniz@yahoo.com.br}
\author{D. Z. Rossatto}
\email{zini@df.ufscar.br}
\author{C. J. Villas-Boas}
\email{villasboas@ufscar.br}
\affiliation{Departamento de F\'{i}sica, Universidade Federal de S\~{a}o Carlos, 13565-905 S\~{a}o Carlos, S\~{a}o Paulo, Brazil}

\begin{abstract}
We theoretically investigate the implementation of the two-mode squeezing operator in circuit quantum electrodynamics. Inspired by a previous scheme for optical cavities [Phys. Rev. A \textbf{73}, 043803(2006)], we employ a superconducting qubit coupled to two nondegenerate quantum modes and use a driving field on the qubit to adequately control the resonator-qubit interaction. Based on the generation of two-mode squeezed vacuum states, firstly we analyze the validity of our model in the ideal situation and then we investigate the influence of the dissipation mechanisms on the generation of the two-mode squeezing operation, namely the qubit and resonator mode decays and qubit dephasing. We show that our scheme allows the generation of highly squeezed states even with the state-of-the-art parameters, leading to a theoretical prediction of more than 10 dB of two-mode squeezing. Furthermore, our protocol is able to squeeze an arbitrary initial state of the resonators, which makes our scheme attractive for future applications in continuous-variable quantum information processing and quantum metrology in the realm of circuit quantum electrodynamics.

\end{abstract}


\maketitle

\section{Introduction}

Squeezed states of electromagnetic fields are characterized by presenting uncertainty in the fluctuations of one of their quadratures smaller than that expected for coherent states, at expenses of an increase in the fluctuations of the conjugated quadrature \cite{Walls}. Among other applications, such kind of states, initially proposed by C. Caves \cite{Caves1981}, have been recently employed in the improving of high precision quantum measurements, such as in the detection of gravitational waves \cite{BRSLIGO}. Moreover, an special class of squeezed states, namely two-mode squeezed vacuum states (TMSSs), besides being a resource for quantum metrology \cite{Dow2010}, is a cornerstone for many quantum information processing tasks such as quantum teleportation \cite{Loock}, since the TMSS  is the quantum optical representative for bipartite continuous-variable entanglement.

Thus, the generation of such states with high degree of squeezing is a subject of continuous investigation.To mention some recent advances, experimental generation as well as theoretical proposals of generation of  single- \cite{Beltran,zag2008,Mallet,zag2012,zag2012-2} and two-mode \cite{xue2007,joh2009,joh2010,Wil,C,F,M,wal2014} squeezed states have been reported using superconducting circuits \cite{Gu2017}. Also, some circuit quantum electrodynamics (QED) experiments involving the interaction of artificial atoms with squeezed electromagnetic vacuum have confirmed theoretical predictions of the 1980s in the context of quantum optics \cite{nor2011,MT}. In addition, some schemes that use parametric down conversion process to generate squeezed states of one or two modes have been investigated in the context of both cavity QED \cite{RMCVP,Fabiano2006} and circuit QED \cite{K,W,ZGA}.

Here we extend the protocol for the implementation of the two-mode squeezing operator in optical cavities \cite{Fabiano2006} to the context of a circuit-QED setup, introducing a detailed analysis of the validity of the employed approximations and of the influence of the dissipative mechanisms on the fidelity of the squeezing process. With the present protocol one would be able to squeeze an arbitrary two-mode ($a$ and $b$) initial state $|\Psi(0)\rangle_{ab}$, i.e., here we show how to implement the operation $S(\zeta)_{ab}|\Psi(0)\rangle_{ab}$, being $S(\zeta)_{ab}$ the two-mode squeezing operator with squeezing parameter $\zeta$ \cite{Walls}. Although an arbitrary two-mode initial state can be squeezed, here we focus on the generation of TMSSs to perform the aforementioned analysis. In Ref. \cite{Fabiano2006} it is proposed different schemes for the generation of single- and two-mode squeezed states in optical cavities, only the latter can be extended to the context of circuit QED since the protocol used for the generation of single mode states requires two intense classical fields driving the qubit, being one of them so intense that would not validate the approximations performed in circuit QED.

We show that, with the current technology, our scheme is able to generate high degrees of squeezing in circuit QED. To this end, we must consider a superconducting flux qubit (artificial atom) \cite{orlando,Bylander2011} dispersively coupled to two spatially separated resonators. To engineer the desired Hamiltonian, the qubit must be resonantly driven by a single classical (external) field. Even though other artificial atoms, as the transmon qubit, can present longer coherence times, as recently demonstrated in 3D superconducting architectures \cite{PR,rigetti2012}, they are not suitable for our protocol. This happens because their energy-level anharmonicities are not strong enough to avoid unwanted transitions induced by the strong driving field required to engineer the desired effective interaction. 

In this paper we firstly analyze the validity of the approximations employed in the derivation of the effective Hamiltonian, investigating the ideal scenario for the implementation of the two-mode squeezing operator in circuit QED. Differently from Ref.~\cite{Fabiano2006}, we provide a more detailed study of the validity of the approximations employed in
the derivation of the effective Hamiltonian, and also perform a complete study regarding the influence of the qubit and resonator dissipative processes on the generation of TMSSs. We note that, as our scheme requires superposition states for the qubit, its decay and dephasing rates become the critical parameters. Nevertheless, our scheme predicts high degrees of squeezing even for the parameters achieved in present-day circuit-QED experiments.

\section{The model} 

Considering a single artificial atom up to its second excited state (flux qubit plus a third level), the dynamics of it interacting with two resonators and an external classic field can be written as
\begin{align}
H_{\text{full}}&=\omega _{a}a^{\dag }a +\omega _{b}b^{\dag }b +\frac{\omega _{0}}{2}\sigma_{z} + \left(\frac{\omega _{0}}{2}+\omega_{\text{ef}}\right)\sigma_{\text{ff}} \nonumber \\
&+\left[g_{a}( a+ a^{\dagger}) + g_{b}( b+ b^{\dagger})+2\Omega \cos{(\omega_{d}t)}\right] \nonumber \\
&\times [(\sigma _{\text{ge}}+\sigma _{\text{eg}}) + (\sigma _{\text{ef}}+\sigma _{\text{fe}})], \label{eq:full}
\end{align}  
in which $a$ ($a^{\dagger}$) and $b$ ($b^{\dagger}$) are the annihilation (creation) operators for the respective modes, which have frequencies $%
\omega _{a}$ and $\omega _{b}$. Here, $\sigma_{z} = |\text{e}\rangle \langle \text{e}|-|\text{g}\rangle \langle \text{g}|$ and $\sigma_{\text{eg}} = \sigma_{\text{ge}}^{\dagger} = |\text{e}\rangle \langle \text{g}|$ are the qubit operators, being $|\text{e} \rangle$ and $|\text{g} \rangle$ the excited and ground states of the qubit, respectively, while $\sigma_\text{ff} = |\text{f}\rangle \langle \text{f}|$ and $\sigma_{\text{ef}} = \sigma_{\text{fe}}^{\dagger} = |\text{e}\rangle \langle \text{f}|$ with $|\text{f} \rangle$ being the second excited state of the artificial atom. The transition frequency between  $|\text{g} \rangle$ and  $|\text{e} \rangle$ is $\omega _{0}$ while the one between $|\text{e} \rangle$ and  $|\text{f} \rangle$ is $\omega _\text{ef}$. The qubit-mode ``$a$" (``$b$") coupling is represented by $g_a$ ($g_b$). Finally, the flux qubit is driven by a resonant external classical field, being $2\Omega$ its Rabi frequency while $\omega _{d}$ its oscillation frequency. The modes and the external driving field also couple the transition $|\text{e} \rangle \leftrightarrow |\text{f} \rangle$, and for the sake of simplicity we consider the same coupling strengths $g_{a}$ ($g_{b}$) and $2\Omega$.

When $\omega_{y} + \omega_{x} \gg |\omega_{y} - \omega_{x}| \gg |g_{x}|$, with $y=\{0,\text{ef}\}$ and $x=\{a,b,d\}$ ($g_{d} \equiv 2\Omega$), we can neglect the counter-rotating terms (rotating-wave approximation) in Eq.~(\refeq{eq:full}). Moreover, for a qubit with very large anhamonicity ($|\omega_{ef}-\omega_{0}|/\omega_{0} \gg 1$), its third level does not affect substantially the dynamics and can be neglect without loss of generality. With these considerations the Hamiltonian can be reduced to \cite{MA}
\begin{equation}
H = H_{0} + H_{I}, 
\label{eq:e1}
\end{equation}
with
\begin{align}
&H_{0}=\omega _{a}a^{\dag }a +\omega _{b}b^{\dag }b +\frac{\omega _{0}}{2}\sigma_{z}, \label{eq:e2} \\
&H_{I}=\left( g_{a}a+g_{b}b +\Omega e^{-i\omega_{d}t}\right)\sigma _{\text{eg}}+ \text{H.c.} \label{eq:e3}
\end{align}  

\begin{figure}[t]
\centering
\includegraphics[trim = 40mm 95mm 170mm 10mm, clip, width=0.35\textwidth]{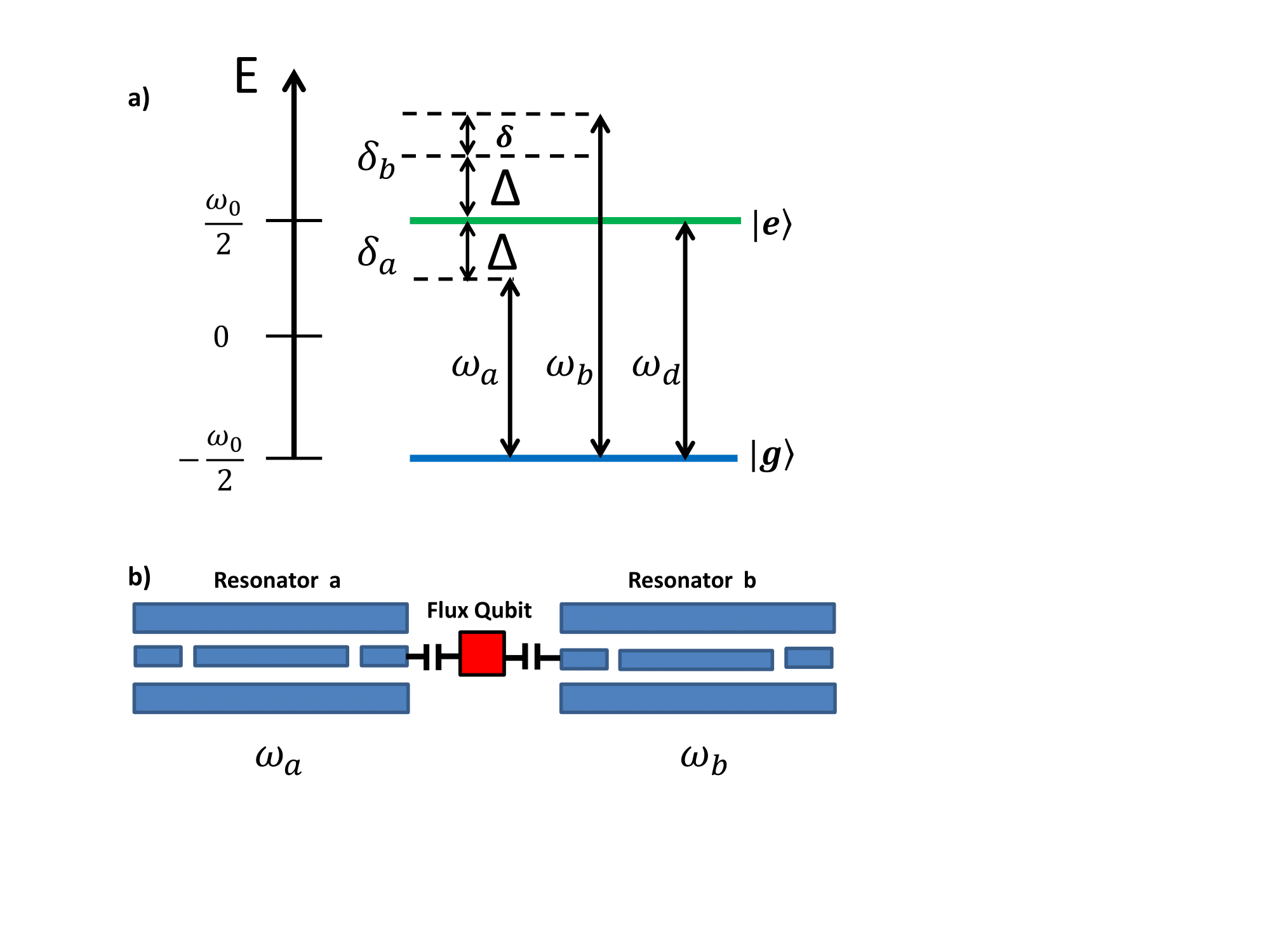}
\caption{(a) Energy-level diagram of the qubit and relevant frequencies involved in our scheme. (b) Pictorial representation of the circuit-QED setup used to generate TMSS, similar to the one employed in Ref. \cite{MA}, which is composed by two resonators coupled to a flux qubit.}
\label{fig:f1}
\end{figure}

In Fig.~\ref{fig:f1}(a) we have the energy-level configuration of the qubit with the relevant frequencies involved, with $\delta_a = \omega_a - \omega_0$, $\delta_b = \omega_b - \omega_0$, and $\omega_d = \omega_0$. In Fig.~\ref{fig:f1}(b) we show a pictorial representation of our circuit-QED setup. This configuration is similar to that employed in Ref.~\cite{MA} to study the generation of TMSSs through reservoir engineering. While our scheme requires a monochromatic microwave field that transversely drives the qubit, the protocol in Ref.~\cite{MA} requires that the qubit is longitudinally driven by a bichromatic microwave field. Furthermore, the protocol presented in Ref.~\cite{MA} has the advantage of generating a TMSS as a stationary state and thus being robust against the decoherence, but it cannot be used to squeeze an arbitrary initial state as ours.  A similar experimental setup is also used in \cite{Wang2016} to experimentally generate a Schr\"{o}dinger two-mode cat state.

Rewriting $H$ in the interaction picture, we have
\begin{equation}
V_{I}(t)= \left(g_{a}e^{-i\delta _{a}t}a+g_{b}e^{-i\delta_{b}t}b+\Omega\right) \sigma _{\text{eg}}+ \text{H.c.}
\label{eq:e4}
\end{equation}
Applying a second unitary transformation given by $U=\exp \left[-i\left( \Omega \protect\sigma_{\text{eg}}+\Omega ^{\ast }\protect\sigma _{\text{ge}}\right) t\right] $, we end up with the Hamiltonian $\mathcal{V}_{I}(t)=U^{-1}V_{I}(t)U-\left( \Omega \sigma _{\text{eg}}+\Omega^{\ast }\sigma _{\text{ge}}\right)$, which contains only highly oscillating terms, allowing us to apply the method employed in Ref.~\cite{James2000} to obtain the effective Hamiltonian for our system. Namely, considering up to second-order processes, $H_\text{eff}=-i\mathcal{V}_{I}(t) \int \mathcal{V}_{I}(t)dt$.

For $\delta_a = \Delta$ and $\delta_b = -\Delta -\delta$ ($|\delta| \ll |\Delta|$), under the large detuning and strong driving field conditions, i.e., $|\Delta +2\Omega |\gg |\eta |$ with $\eta =2\Omega -\Delta $, and also considering $|\eta| \gg g$, we derive the effective Hamiltonian within the rotating-wave approximation (neglecting the residual highly oscillating terms)  \cite{Fabiano2006}
\begin{align}
H_\text{eff}=&\left( \chi _{a}aa^{\dagger }+\chi _{b}b^{\dagger }b\right) \sigma_{++}-\left( \chi _{a}a^{\dagger }a+\chi _{b}bb^{\dagger }\right) \sigma_{--} \nonumber \\
&+\left( \frac{g_{a}g_{b}e^{i\delta t}}{4\eta }ab+ \text{H.c.}\right) \left(\sigma _{--}-\sigma _{++}\right),
\end{align}
in which $\chi _{\alpha }=|g_{\alpha }|^{2}/4\eta $ ($\alpha =a,b$) and $\sigma_{\pm} = |\pm \rangle \langle \pm|$, with $\left\vert \pm \right\rangle =\frac{1}{\sqrt{2}}\left( \left\vert \text{g}\right\rangle \pm \left\vert \text{e}\right\rangle \right) $. 

From the effective Hamiltonian above, we immediately see that the qubit states $|+\rangle$ and $|-\rangle$ give rise to independent dynamics. For both qubit states we can adjust the detuning $\delta$ to obtain efficient squeezing processes. For instance, let us consider the system initially in the state $\left\vert \psi \left( 0\right) \right\rangle =\left\vert -\right\rangle _{\text{atom}}\otimes $ $\left\vert \Psi(0)\right \rangle_{ab} $. Applying the unitary transformation $U_{-}=\exp \left( i \chi_a a^{\dagger} a t + i \chi_b b^{\dagger} bt \right)$, the effective Hamiltonian for the resonators can be reduced to 
\begin{equation}
\mathcal{H}_{-} = \frac{g_{a} g_{b} }{4\eta }ab+\text{H.c.}
\label{eq:e7}
\end{equation} 
if we adjust $\delta =-\left( \chi _{a}+\chi _{b} \right) $. This Hamiltonian is exactly the one that allows the generation of ideal TMSSs \cite{Walls}, being $\lambda \equiv \left( \frac{g_{a} g_{b}}{4\eta } \right)$ the effective coupling constant. In the ideal case, the evolution of this system is simply given by $\left\vert \psi \left( \tau \right) \right\rangle = \exp\left(-i \lambda\tau ab - \text{H.c} \right)|\Psi(0)\rangle_{ab} =  S(\zeta)_{ab}|\Psi(0)\rangle_{ab}$, with $\zeta = -i\lambda \tau$ while $\tau$ represents the interaction time between the qubit and the modes. Thus, this scheme allows us to squeeze an arbitrary initial two-mode state with the squeezing factor given by $r = |\lambda| \tau$.

\section{Analyses of the validity of the approximations}

\subsection{Unitary dynamics}  

Firstly we have to investigate the validity of our approximations carried out above. As our scheme requires a strong driving field and non-resonant interactions, we must be sure about the range of validity of the parameters. To this end we compare the dynamics of the effective Hamiltonian $\mathcal{H}_{-}$ [Eq. (\ref{eq:e7})] with $V_{I}(t)$ [Eq.(\ref{eq:e4})] and also with the Hamiltonian without any approximation $H_\text{full}$ [Eq.~(\ref{eq:full})]. For this comparison we focus on the generation of a TMSS. To quantify the degree of squeezing, we employ the total variance of EPR-like operators  \cite{DUAN} 
\begin{equation}
\text{V}_{\text{ar}}=\langle \left( \Delta u \right)^{2}+\left( \Delta v \right) ^{2}\rangle ,
\end{equation}
in which $u= X_{a}+ X_{b}$ and $v=P_{a}-P_{b}$, where the position and momentum quadrature operators are defined as $X_{\alpha}=\left( \alpha e^{-i\theta}+\alpha^{\dag }e^{i\theta}\right)  /\sqrt{2}$ and $P_{\alpha}=-i\left( \alpha e^{-i\theta}-\alpha^{\dag }e^{i\theta}\right) /\sqrt{2},$ with $\left( \alpha = a, b \right) $, respectively. The parameter $\theta$ refers to the squeezing direction, which depends on the phases of the driving field and qubit-mode couplings. This variance is an important quantifier of the degree of squeezing of two-mode states and works out as a witness of entanglement (a two-mode state is entangled whenever $\text{V}_{\text{ar}}<2$) \cite{DUAN}. In experimental works one usually quantifies the degree of squeezing in decibels (dB), which is connected to the definition above via the expression $-10\log_{10}(V_{\text{ar}}/2)$ \cite{KJG}.

\begin{figure}[t]
\includegraphics[trim = 5mm 0mm 10mm 10mm, clip, width=0.33\textwidth]{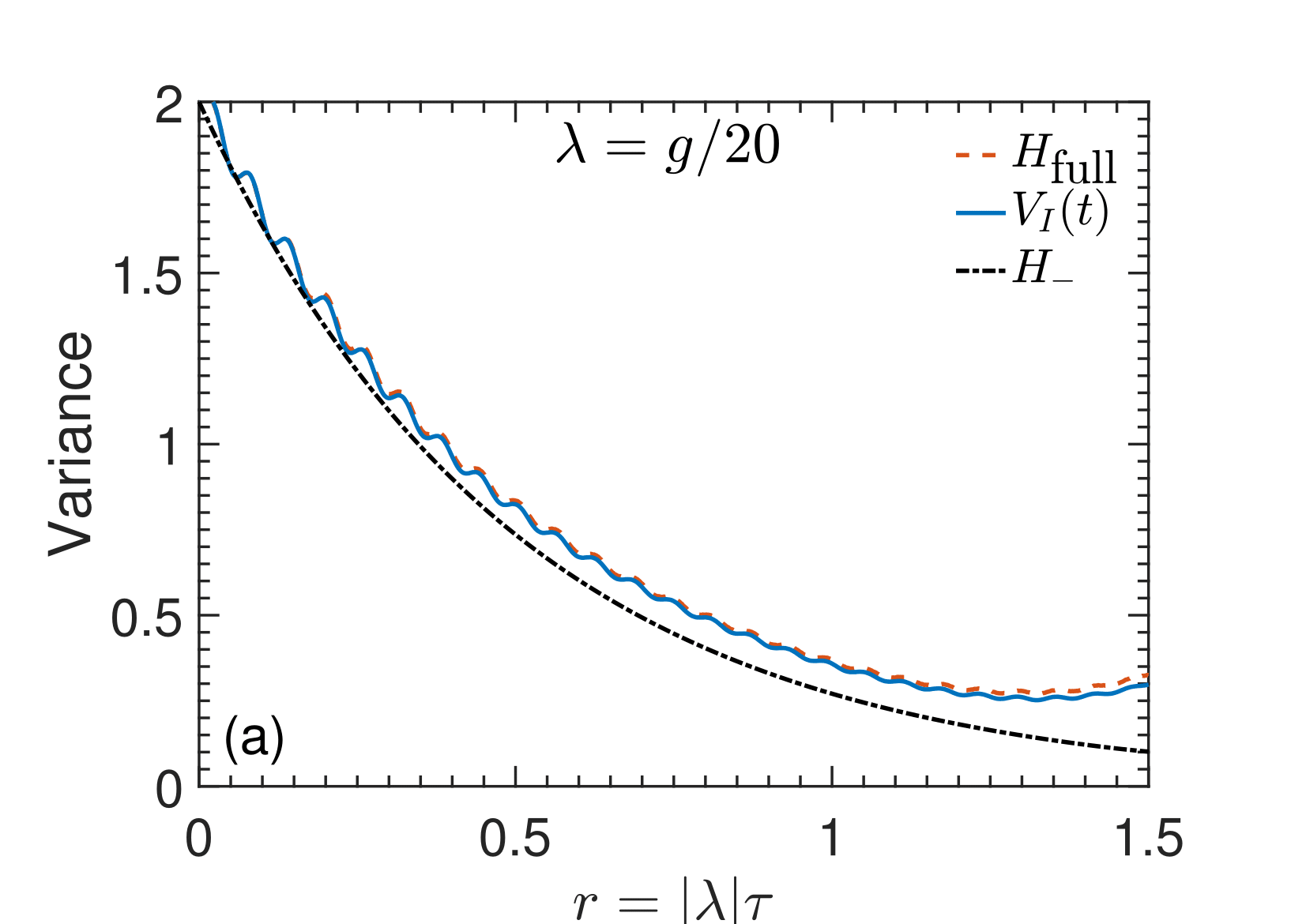}
\includegraphics[trim = 5mm 0mm 10mm 10mm, clip, width=0.33\textwidth]{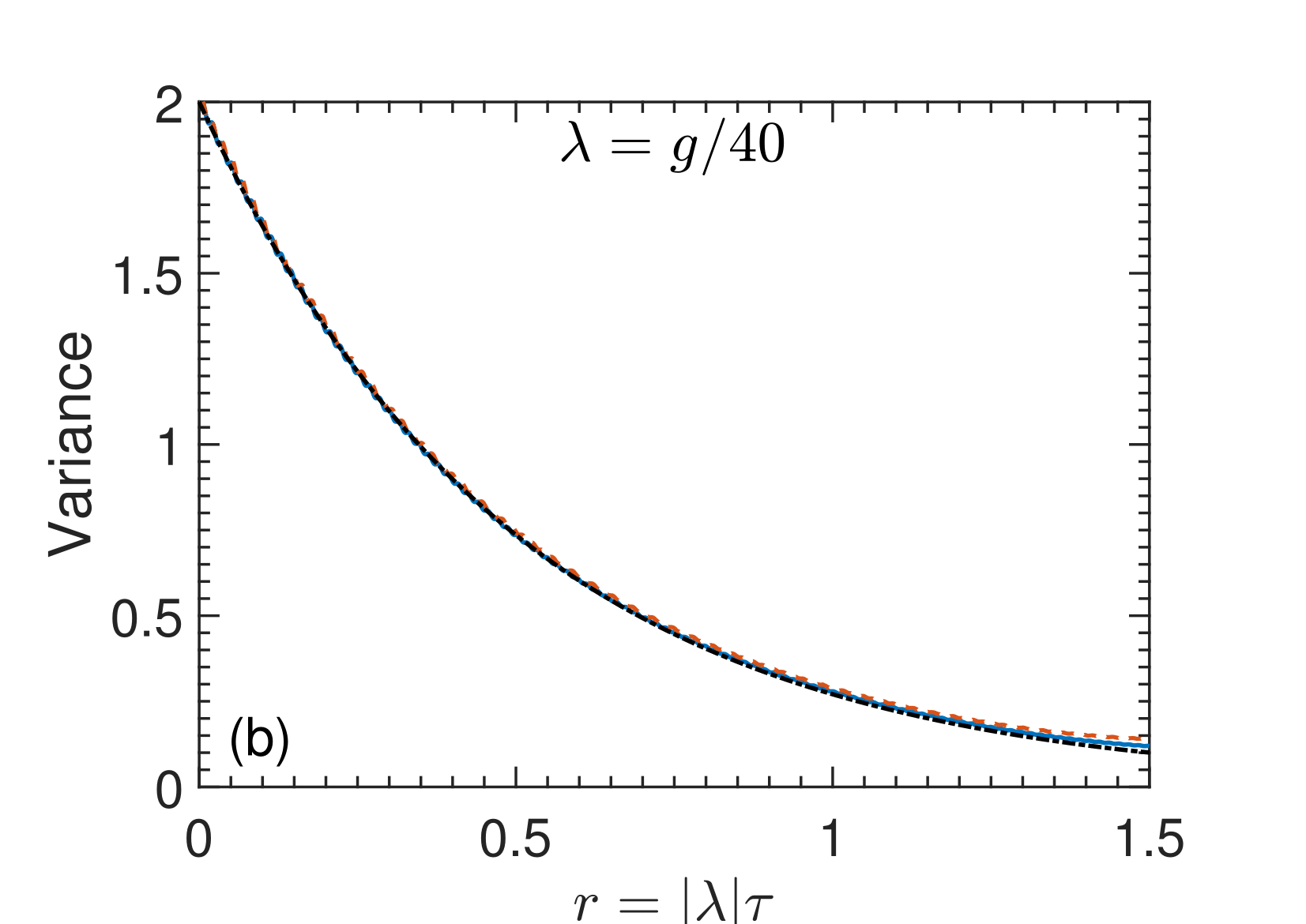}
\includegraphics[trim = 5mm 0mm 10mm 10mm, clip, width=0.33\textwidth]{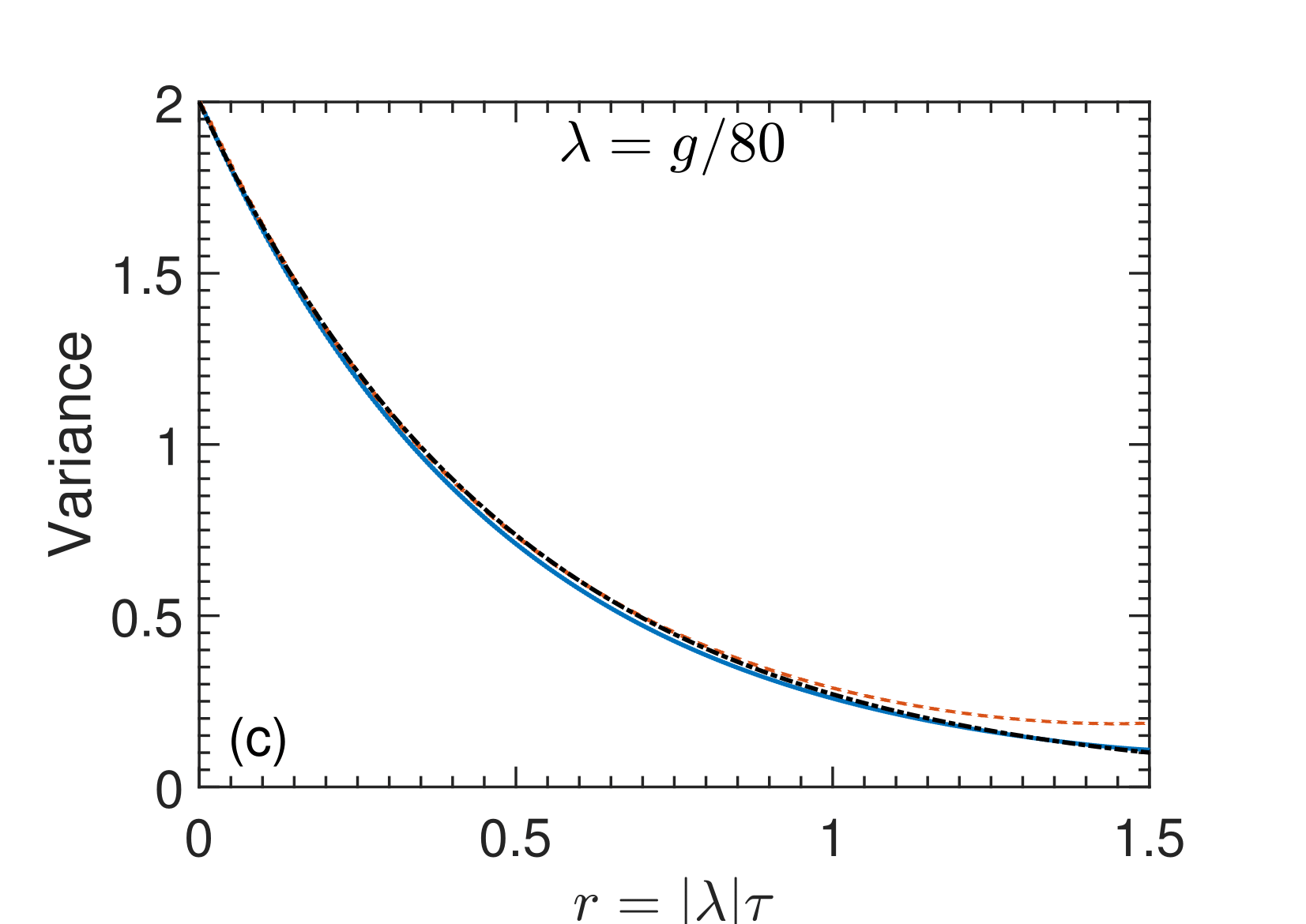}
\caption{Variance as a function of $r = \vert\lambda\vert \tau$ for different parameter regimes, comparing $H_\text{full}$ [Eq. (\ref{eq:full})] (red dashed line), $V_{I}(t)$ [Eq.~(\ref{eq:e1})] (blue solid line) and $H_{-}$ [Eq. (\ref{eq:e7})] (black dot-dashed line). We have fixed $g_a = g_b = g$, $\omega_{0} = 500g$, $\omega_\text{ef} = 5\omega_{0}$, $\theta =\pi/4$ (which is the squeezing axis for the parameters used here) and consider (a) $\Delta = 35 g$ and $\Omega = 20 g$ ($\lambda=g/20$), (b) $\Delta = 90 g$ and $\Omega = 50 g$ ($\lambda=g/40$), and (c) $\Delta = 180 g$ and $\Omega = 100 g$ ($\lambda=g/80$). In all curves we have neglected the dissipative processes.}
\label{fig:f2}
\end{figure} 

In Fig.~\ref{fig:f2} we plot $\text{V}_{\text{ar}}$ as a function of the ideal squeezing factor, $r = \vert\lambda\vert \tau$, for different parameter regimes assuming the modes initially in the vacuum state. We clearly see that, increasing $|\Delta|$, $|\Omega|$, and $|\eta|$, i.e., decreasing $|\lambda|$, the dynamics given by $V_{I}(t)$ approaches better and better the desired one. The smaller $\lambda$ the higher the degree of squeezing allowed to be reached, but the longer the interaction time required, and then the dissipative processes can also play an important role as shown later.


We also have to be careful with too strong $|\eta|$ (high values of $\Delta$ and $\Omega$), since in this limit the model given by Eq.~\eqref{eq:e1} can become no longer valid as we can see in the last panel of Fig.~\ref{fig:f2}. For instance, assuming the reasonable value of the qubit-field coupling $g/2\pi = 20$ MHz would imply $\Omega/2\pi = 2.0$ GHz for the values used for $\lambda = g/80$, which is a value at which the rotating-wave approximation starts to fail (for qubit-driving field interactions at least) \cite{Yoshihara2014}, and therefore $H_\text{full}$ can no longer be well described by $V_{I}(t)$.

This value of the Rabi frequency of the classical field would also induce transitions to other levels in transmon qubits, as their energy-level anharmonicities are not strong enough: the transition frequency from the first excited state to the ground  $\omega_{0}/2\pi = 8.6$ GHz is close to that from the second excited state to the first one (i.e., $\omega_\text{ef}-\omega_0 = - 421$MHz), implying on a ratio of $|\omega_\text{ef}|/\omega_{0}= 0.95$ only \cite{Pechal2014}. Thus, a classical field with Rabi frequency  $\Omega/2\pi = 2.0$ GHz, addressing the ground to first excited state transition, would certainly induce transitions from the first to the second excited state and then our effective model would be no longer valid. This fact prohibits us considering transmon qubits in our protocol. However, flux qubits have stronger anharmonicities, providing the ratio $|\omega_\text{ef}|/ \omega_0 = 5$ \cite{Bylander2011} and then, even for $\Omega/2\pi = 2.0$ GHz, we are able to neglect the transitions to higher excited states induced by the classical field, but in this case we have to be careful with the validity of the rotating-wave approximation in the qubit-driving field Hamiltonian.

In the following we analyze the influence of both resonator and qubit losses on the process of generation of squeezed states. For that we consider $\lambda = g/40$, a parameter with which we have seen that $V_{I}(t)$ provides a dynamics in excellent agreement with the Hamiltonian without any approximation ($H_\text{full}$).

\subsection{Dissipative Dynamics} 
If the qubit and the resonators are coupled to their respective reservoirs under the Born-Markov approximation, we can take into account the dissipative effects on our system by using the master equation in the Lindblad form \cite{Carmichael} 
\begin{equation}
\dot{\rho}=-i[V_{I}(t),\rho] +\frac{\gamma }{2}\mathcal{D}[ \sigma _{-}]\rho +\frac{\gamma_{\text{ph}}}{2}\mathcal{D}[ \sigma _{ee}]\rho + \sum_{\alpha=a,b}\frac{\kappa _{\alpha}}{2}\mathcal{D}[\alpha] \rho,
\label{eq:master_equation} 
\end{equation}
with $\mathcal{D}[ \mathcal{O}]\rho = 2\mathcal{O}\rho \mathcal{O}^{\dagger} - \mathcal{O}^{\dagger}\mathcal{O}\rho - \rho\mathcal{O}^{\dagger}\mathcal{O} $.
The first term describes the unitary evolution while the last ones describe the dissipation on the qubit (decay rate $\gamma$), its dephasing (rate $\gamma_{\text{ph}}$), and on the resonators (decay rate $\kappa_\alpha$), respectively. We are allowed to use this master equation instead of the dressed one \cite{Beaudoin2011} since we are assuming neither the ultrastrong nor the deep strong coupling regimes, i.e., our results are valid for $g_\alpha/\max{(\omega_{0},\omega_{\alpha})} < 0.1$ ($\alpha = a,b$). Even below this limit, one knows that dispersive interaction can introduce corrections onto the standard master equation proportional to $(g_\alpha/\Delta)^2$ and the mean number of photons in the resonators \cite{Boi2009}. However, as our protocol requires $|\Delta| \gg g_\alpha$ and we deal with not so high mean number of photons, we can neglect such corrections to the master equation above. Due to the high dimension of the Hilbert space of our system, we numerically solve the master equation using the Monte Carlo wave-function method \cite{Carmichael}, with the help of the QuTIP algorithm \cite{JNATION}.

\begin{figure}[t]
\centering
\includegraphics[trim = 0mm 0mm 8mm 10mm, clip, width=0.24\textwidth]{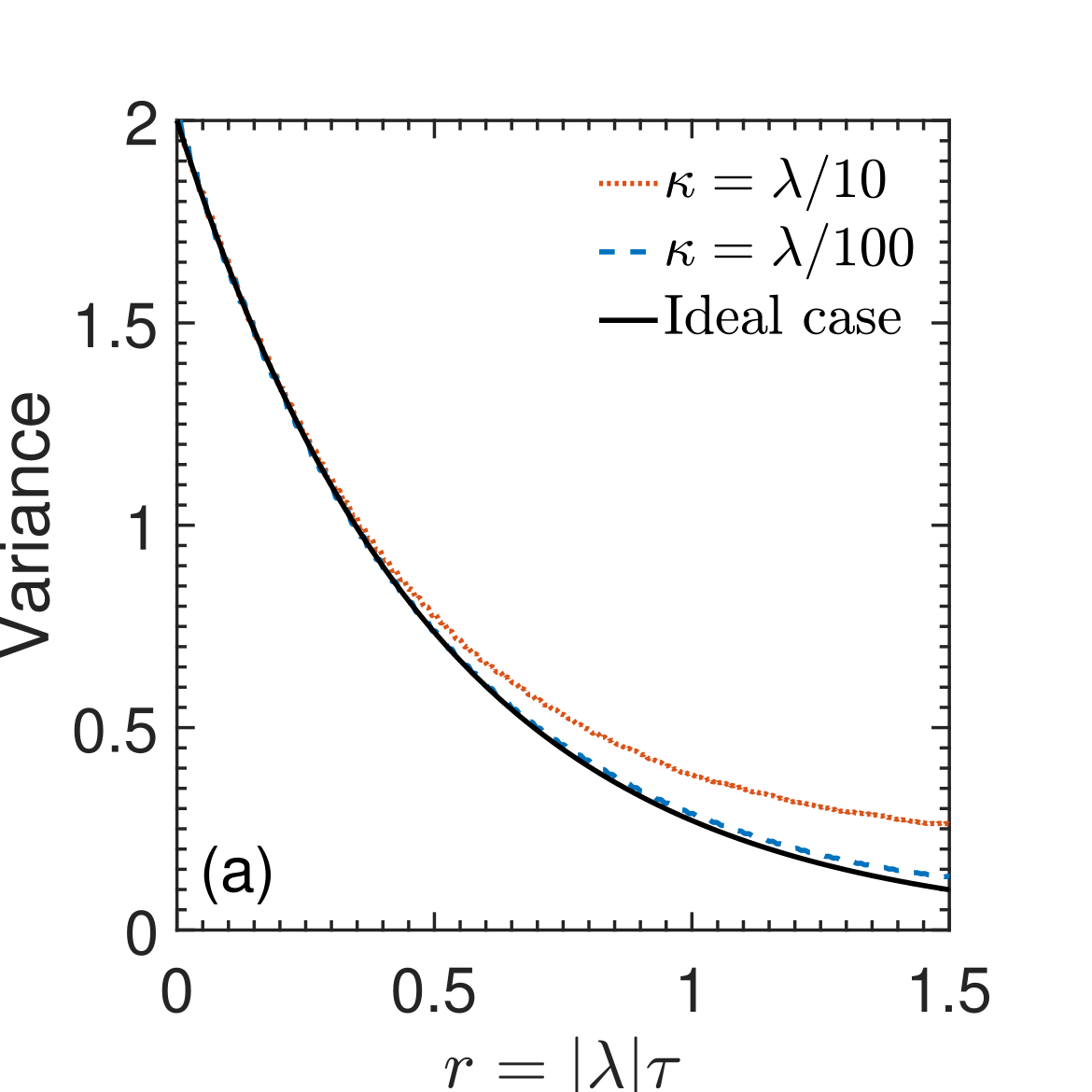}
\includegraphics[trim = 0mm 0mm 8mm 10mm, clip, width=0.24\textwidth]{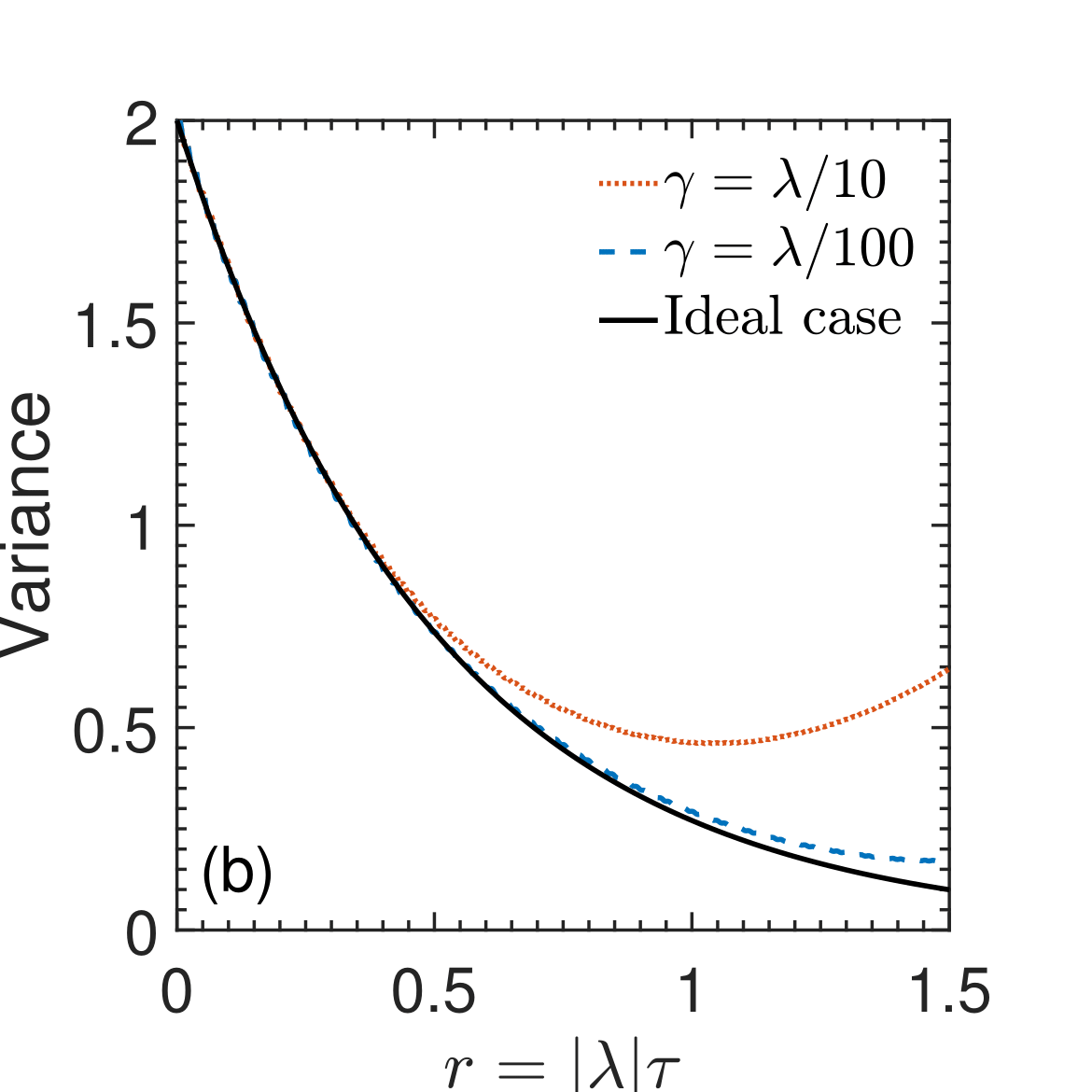}
\includegraphics[trim = 0mm 0mm 8mm 10mm, clip, width=0.24\textwidth]{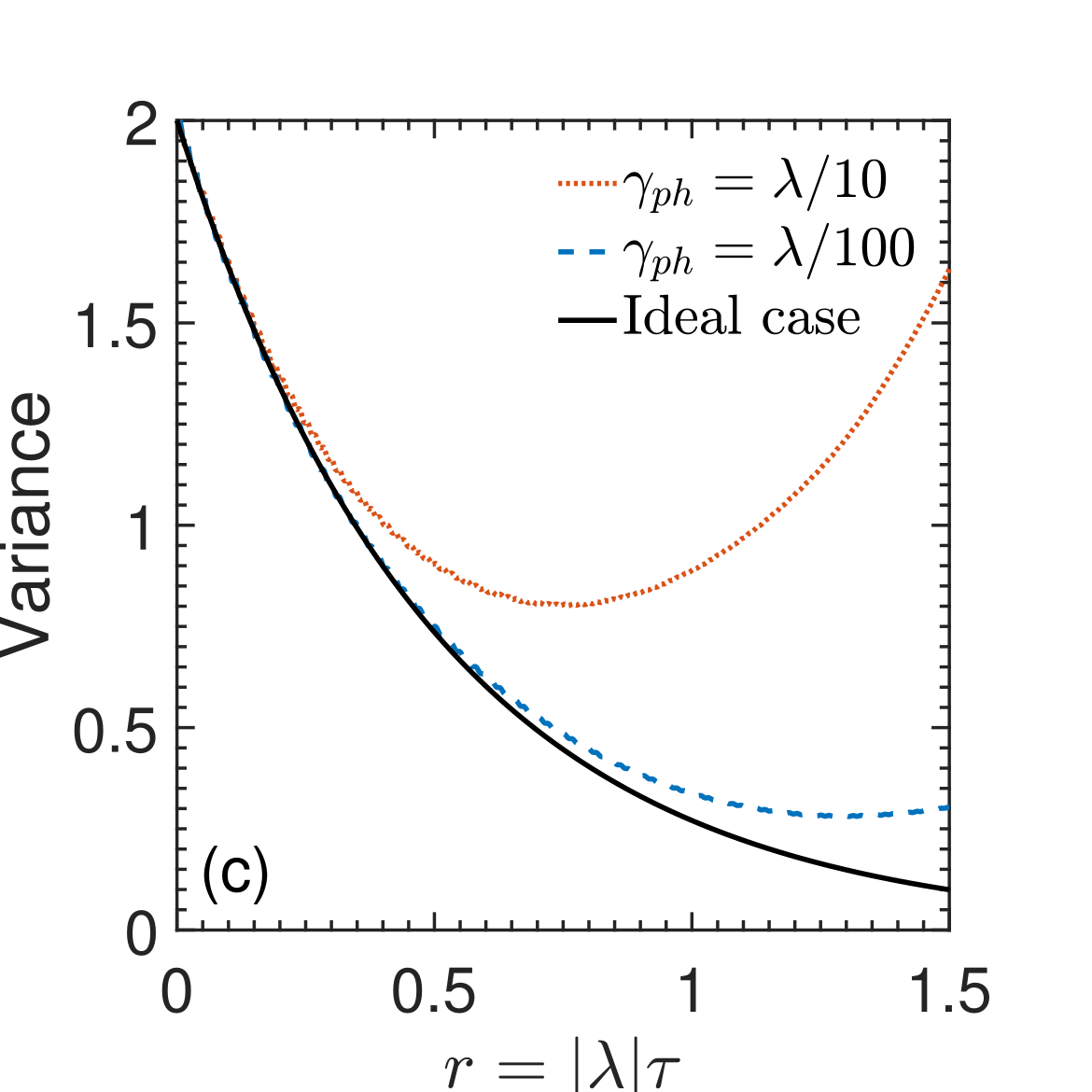}
\includegraphics[trim = 0mm 0mm 8mm 10mm, clip, width=0.24\textwidth]{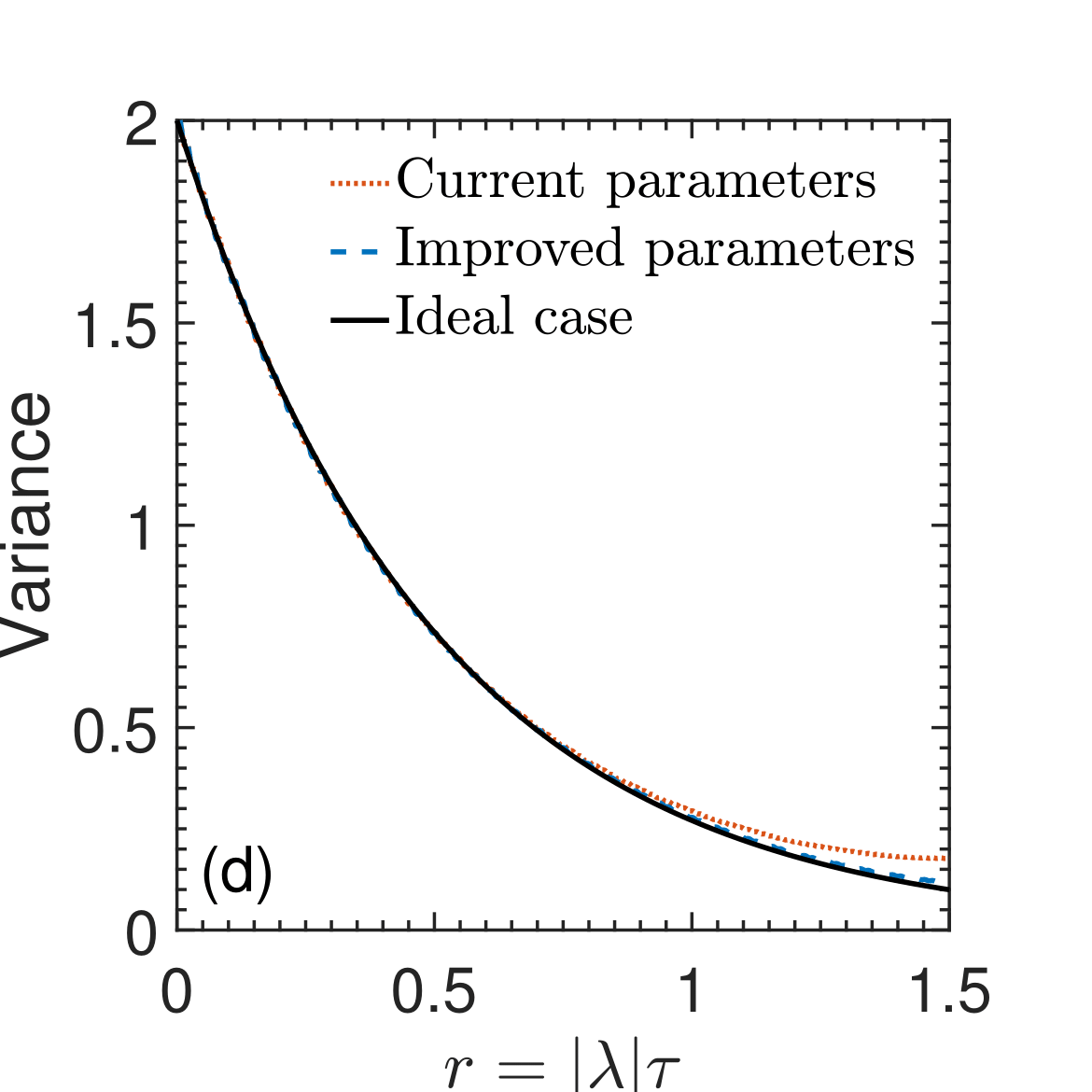}
\caption{Influence of the dissipation and decoherence processes on the generation of TMSS. In all plots we show the Variance, $\text{V}_{\text{ar}}$, as a function of $r = \vert\lambda\vert \tau$ for $\Delta = 90 g$ and $\Omega = 50 g$, i.e., $\lambda = g/40$. The full black lines represent the ideal squeezing process. (a) Without qubit dissipation and dephasing ($\gamma = \gamma_\text{ph} = 0$). (b) Ideal resonators ($\kappa = 0$) and without qubit dephasing ($\gamma_\text{ph}=0$). (c) Ideal resonators ($\kappa = 0$) and without qubit dissipation ($\gamma=0$). In (d) we plot the evolution of the system under the action of all reservoirs for present-day parameters (dotted red line). We have assumed $g = 2\pi\times20$ MHz, which gives $\lambda = 2\pi\times500$ KHz. Thus, $\gamma = T_{1}^{-1}\simeq 5.2 \times 10^{-3} \lambda$ and $\gamma_\text{ph} =T_{2}^{*-1}-(2T_{1})^{-1} \simeq  1.0 \times 10^{-3} \lambda$ ($T_1 = 60\mu$s and a coherence time $T_{2}^{*} = 85\mu$s \cite{Yan2016}), and $\kappa =\kappa_a = \kappa_b \simeq 2\pi\times 53$ Hz ($\kappa^{-1} \simeq 3$ ms \cite{Wang2016}), which results in $\kappa \simeq 1.0\times10^{-4}\lambda$. An almost ideal squeezing process can be achieved if we are able to improve in one order of magnitude the current energy relaxation and coherence times of the flux qubit (i.e., we have assumed $\gamma=\gamma_\text{real}/10$ and $\gamma_\text{ph}=\gamma_\text{ph}^\text{real}/10$) as we see in the dashed blue line.}
\label{fig:f3new}
\end{figure} 

\textit{Resonator dissipation.---} To understand the role of each dissipation channel, firstly we analyze the influence of the dissipation on the resonators, which we are assuming identical for the sake of simplicity, i.e., $\kappa_a = \kappa_b = \kappa \neq 0$.
In Fig.~\ref{fig:f3new}(a) we plot $\text{V}_{\text{ar}}$ as a function of the ideal squeezing parameter ($r$), considering the parameters that results in $\lambda=g/40$, without dissipation and dephasing on the qubit ($\gamma =\gamma_\text{ph} = 0$), and different values for the dissipation on the resonators. The dissipation on the modes introduces a competition: the interaction with the ideal qubit squeezes the two-mode field while the resonator decays disentangle pairs of correlated photons, destroying the squeezing. Therefore, the bigger the ratio $\vert\lambda\vert/\kappa$, the higher the achieved degree of squeezing, as shown in Fig. \ref{fig:f3new}(a).

\textit{Qubit dissipation.---} Now we consider the dissipation on the qubit only, that is, $\gamma \neq 0$ and $\gamma_\text{ph} = \kappa_a = \kappa_b = 0$). Here the interaction of the qubit with the environment will destroy the initial atomic superposition, assumed as $|-\rangle = \left( |\text{g}\rangle - |\text{e}\rangle \right)/\sqrt{2}$. Thus, the dissipation on the qubit will drive it to a different state, projecting the effective Hamiltonian in other different from $\mathcal{H}_{-}$. For instance, when the qubit decays, we end up with the qubit state $|\text{g}\rangle = \left( |+\rangle + |-\rangle \right)/\sqrt{2}$ and then the effective dynamics would be given by a mixture of $\mathcal{H}_{-}$ (state $|-\rangle$) and $\mathcal{H}_{+}$ (state $|+\rangle$), which squeezes the cavity modes in orthogonal directions. Hence, the mixture of these two squeezing process (in orthogonal directions) no longer generates an ideal TMSS. This fact can be clearly seen in Fig.~\ref{fig:f3new}(b). For long interaction times, the qubit decay plays a more prominent negative role than the resonator decays in the squeezing process.

\textit{Qubit dephasing.---} Here we analyze the role of the qubit dephasing on the generation of two-mode squeezed states, that is, $\gamma_\text{ph} \neq 0$ and $\gamma = \kappa_a = \kappa_b = 0$, as we see in Fig. \ref{fig:f3new}(c). As happened in the previous case, the dephasing process will also destroy the initial atomic superposition and then this process will also greatly damage the generation of highly squeezed states. In fact, we see that this decoherence channel is the one that most damages the generation of TMSS.

\textit{Resonator and qubit dissipation.---} Finally we investigate a real scenario, taking into account the dissipation on both resonators (assumed equal) and the dephasing and decaying of the flux qubit [see Fig. \ref{fig:f3new}(d)]. Again, we fixed $\Delta = 90 g$ and $\Omega = 50 g$ ($\lambda = g/40$), considering different values for the decay rates. In particular, we consider the state-of-the-art parameters (blue dashed line) based on the work by F. Yan \textit{et al.~}\cite{Yan2016}, where the authors report a flux qubit with an energy relaxation time $T_1 = 60\mu$s and a coherence time $T_{2}^{*} = 85\mu$s, such that $\gamma = T_{1}^{-1}\simeq 5.2 \times 10^{-3} \lambda$ and $\gamma_\text{ph} =T_{2}^{*-1}-(2T_{1})^{-1} \simeq  1.0 \times 10^{-3} \lambda$. Considering $\kappa =\kappa_a = \kappa_b \simeq 2\pi\times 53$ Hz ($\kappa^{-1} \simeq 3$ ms) \cite{Wang2016} and assuming $g = 2\pi\times20$ MHz, we have $\lambda = 2\pi\times500$ KHz and $\kappa \simeq 1.0\times10^{-4}\lambda$. For $\vert \lambda \vert \tau \sim 1.5$, such parameters allow a generation of $10.5$ dB of two-mode squeezing ($V_{\text{ar}}\simeq 0.178$), which is very close to the degree recently achieved ($12$ dB) in Ref.~\cite{wal2014}, but the authors use nonlinear resonators while our protocol is based only on linear resonators. On the other hand, better and better degrees of squeezing are possible by enhancing $T_{1}$ and $T_{2}^{*}$. For instance our protocol theoretically predicts a behaviour close the ideal one (up to $r=1.5$, i.e., a two-mode squeezing above $12$ dB) by considering the improvement of one order of magnitude on the present-day relaxation and coherence times $T_{1}$ and $T_{2}^{*}$, respectively [dashed blue line in Fig.~\ref{fig:f3new}(d)].

\section{Conclusions}
We investigated how to implement the two-mode squeezing operator in a circuit-QED system composed by two resonators coupled to a flux qubit, investigating the fidelity of the protocol through its capability of generating highly two-mode squeezed states. The proposed experimental apparatus is essentially the experimental setup employed in Ref.~\cite{Wang2016}. To generate two-mode squeezed states, we must prepare the qubit in the superposition state $|-\rangle$ (the state $|+\rangle$ also allows for the generation of two-mode squeezing operator, requiring only slightly different adjustments). Since our scheme depends on the initial atomic state, the dissipation and the decoherence processes on the qubit will drive the system to a Hamiltonian different from the desired one. Thus, the qubit-resonators interaction time must be shorter than the lifetime (or decoherence) of the qubit. We observe that our scheme theoretically allows for the squeezing of arbitrary initial two-mode states and predicts high degrees of squeezing (more than $10$ dB) with the state-of-the-art parameters, being better with the improvement of such parameters, making the scheme attractive for continuous-variable quantum information processing and quantum metrology.


\begin{acknowledgments}
This work was supported by the S\~{a}o Paulo Research Foundation (FAPESP) Grants No.~2013/04162-5 and 2013/23512-7, the National Council for Scientific and Technological Development (CNPq) Grants No.~161117/2014-7 and 308860/2015-2, and the Brazilian National Institute of Science and Technology for Quantum Information (INCT-IQ) Grant No.~465469/2014-0.

\end{acknowledgments}


\begin{thebibliography}{99}

\bibitem{Walls} Walls, D.F., Milburn, G.J.: Quantum Optics. Springer, Berlin (2008)

\bibitem{Caves1981} Caves, C.M.: Quantum-mechanical noise in an interferometer. Phys. Rev. D \textbf{23}, 1693 (1981)

\bibitem{BRSLIGO}  Schnabel, R.: Squeezed states of light and their applications in laser interferometers. Phys. Rep. {\bf 684}, 1 (2017)

\bibitem{Dow2010} Anisimov, P.M., Raterman, G.M., Chiruvelli, A., Plick, W.N., Huver, S.D., Lee, H., Dowling, J.P.: Quantum metrology with two-mode squeezed vacuum: parity detection beats the heisenberg limit. Phys. Rev. Lett. \textbf{104}, 103602 (2010)

\bibitem{Loock} Braunstein, S.L., van Loock, P.: Quantum information with continuous variables. Rev. Mod. Phys. \textbf{77}, 513 (2005)

\bibitem{Beltran} Castellanos-Beltran, M.A., Irwin, K.D., Hilton, G.C., Vale, L.R., Lehnert, K.W.: Amplification and squeezing of quantum noise with a tunable Josephson metamaterial,. Nat. Phys. \textbf{4}, 929 (2008)


\bibitem{zag2008} Zagoskin, A.M., Il'ichev, E., McCutcheon, M.W., Young, J., Nori, F.: Controlled Generation of Squeezed States of Microwave Radiation in a Superconducting Resonant Circuit, Phys. Rev. Lett. \textbf{101}, 253602 (2008)

\bibitem{Mallet} Mallet, F., Castellanos-Beltran, M.A., Ku, H.S., Glancy, S., Knill, E., Irwin, K.D., Hilton, G.C., Vale, L.R., Lehnert, K.W.: Quantum state tomography of an itinerant squeezed microwave field. Phys. Rev. Lett. \textbf{106}, 220502 (2011)

\bibitem{zag2012} Zagoskin, A.M., Il’ichev, E., Nori, F.: Heat cost of parametric generation of microwave squeezed states, Phys. Rev. A \textbf{85}, 063811 (2012)

\bibitem{zag2012-2} Zagoskin, A.M., Savel'ev, S., Nori, F., Kusmarsev, F.V.: Squeezing as the source of inefficiency in the quantum Otto cycle, Phys. Rev. B \textbf{86}, 014501 (2012)



\bibitem{xue2007} Xue, F., Liu, Y.X., Sun, C.P., Nori, F.: Two-mode squeezed states and entangled states of two mechanical resonators, Phys. Rev. B \textbf{76}, 064305 (2007)

\bibitem{joh2009} Johansson, J.R., Johansson, G., Wilson, C.M., Nori, F.: Dynamical Casimir effect in a superconducting coplanar waveguide, Phys. Rev. Lett. \textbf{103}, 147003 (2009)

\bibitem{joh2010} Johansson, J.R., Johansson, G., Wilson, C.M., Nori, F.: Dynamical Casimir effect in superconducting microwave circuits, Phys. Rev. A \textbf{82}, 052509 (2010)

\bibitem{Wil} Wilson, C.M., Johansson, G., Pourkabirian, A., Simoen, M., Johansson, J.R., Duty, T., Nori, F., Delsing, P.: Observation of the dynamical Casimir effect in a superconducting circuit, Nature (London) \textbf{479}, 367 (2011)

\bibitem{C} Eichler, C., Bozyigit, D., Lang, C., Baur, M., Steffen, L., Fink, J.M., Filipp, S., Wallraff, A.: Observation of two-mode squeezing in the microwave frequency domain, Phys. Rev. Lett. \textbf{107}, 113601 (2011)

\bibitem{F} Flurin, E., Roch, N., Mallet, F., Devoret, M.H., Huard, B.: generating entangled microwave radiation over two transmission lines, Phys. Rev. Lett. \textbf{109}, 183901 (2012)

\bibitem{M} Menzel, E.P., Di Candia, R., Deppe, F., Eder, P., Zhong, L., Ihmig, M., Haeberlein, M., Baust, A., Hoffmann, E., Ballester, D., Inomata, K., Yamamoto, T., Nakamura, Y., Solano, E., Marx. A., Gross, R.: Path entanglement of continuous-variable quantum microwaves, Phys. Rev. Lett. \textbf{109}, 250502 (2012)

\bibitem{wal2014} Eichler, C., Salathe, Y., Mlynek, J., Schmidt, S., Wallraff, A.: Quantum-limited amplification and entanglement in coupled nonlinear resonators, Phys. Rev. Lett. \textbf{113}, 110502 (2014)

\bibitem{Gu2017} Gu, X., Kockum, A.F., Miranowicz, A., Liu, Y.X., Nori, F.: Microwave photonics with superconducting quantum circuits, Physics Reports \textbf{718}, 1 (2017)

\bibitem{nor2011} You, J.Q., Nori, F.: Atomic physics and quantum optics using superconducting circuits, Nature \textbf{474}, 589 (2011)

\bibitem{MT} Murch, K.W., Weber, S.J., Beck, K.M., Ginossar, E., Siddiqi, I.: Reduction of the radiative decay of atomic coherence in squeezed vacuum, Nature (London) \textbf{499}, 62 (2013); Toyli, D.M., Eddins, A.W., Boutin, S., Puri, S., Hover, D., Bolkhovsky, V., Oliver, W.D., Blais, A., Siddiqi, I.: Resonance fluorescence from an artificial atom in squeezed vacuum, Phys. Rev. X \textbf{6}, 031004 (2016)

\bibitem{RMCVP} Villas-Boas, C.J., Moussa, M.H.Y.: One-step generation of high-quality squeezed and EPR states in cavity QED, Eur. Phys. J. D \textbf{32}, 147 (2005)

\bibitem{Fabiano2006} Prado, F.O., de Almeida, N.G., Moussa, M.H.Y., Villas-B\^{o}as, C.J.: Bilinear and quadratic Hamiltonians in two-mode cavity quantum electrodynamics, Phys. Rev. A \textbf{73}, 043803(2006) 

\bibitem{K} Moon, K., Girvin, S.M.: Theory of microwave parametric down-conversion and squeezing using circuit qed, Phys. Rev. Lett. \textbf{95}, 140504 (2005)

\bibitem{W} Wang, Z.H., Sun, C.P., Li, Y.: Microwave degenerate parametric down-conversion with a single cyclic three-level system in a circuit-QED setup, Phys. Rev. A \textbf{91}, 043801 (2015)

\bibitem{ZGA} Zhong, W.-X., Cheng, G.-L., Chen, A.-X.: Int. J. Quantum Inform. \textbf{12}, 1450009 (2014)

\bibitem{orlando} Orlando, T.P., Mooij, J.E., Tian, L., van der Wal, C.H., Levitov, L.S., Lloyd, S., Mazo, J. J.: Superconducting persistent-current qubit, Phys. Rev. B \textbf{60}, 15398 (1999)

\bibitem{Bylander2011} Bylander, J., Gustavsson, S., Yan, F., Yoshihara, F., Harrabi, K., Fitch, G., Cory, D. G., Nakamura, Y., Tsai, J.-S., Oliver, W.D.: Noise spectroscopy through dynamical decoupling with a superconducting flux qubit, Nat. Phys. \textbf{7}, 565 (2011)


\bibitem{PR} Paik, H., Schuster, D.I., Bishop, L.S., Kirchmair, G., Catelani, G., Sears, A.P., Johnson, B.R., Reagor, M.J., Frunzio, L., Glazman, L.I., Girvin, S.M., Devoret, M.H., Schoelkopf, R. J.: Observation of high coherence in josephson junction qubits measured in a three-dimensional circuit qed architecture, Phys. Rev. Lett. \textbf{107}, 240501 (2011)

\bibitem{rigetti2012} Rigetti, C., Gambetta, J.M., Poletto, S., Plourde, B.L.T., Chow, J.M., C\'{o}rcoles, A. D., Smolin, J. A., Merkel, S. T., Rozen, J.R., Keefe, G.A., Rothwell, M.B., Ketchen, M.B., Steffen, M.: Superconducting qubit in a waveguide cavity with a coherence time approaching $0.1$ ms, Phys. Rev. B \textbf{86}, 100506(R) (2012)

\bibitem{MA} Ma, S.-L., Li, Z., Fang, A.-P., Li, P.-B., Gao, S.-Y., Li, F.-L.: Controllable generation of two-mode-entangled states in two-resonator circuit QED with a single gap-tunable superconducting qubit, Phys. Rev. A. \textbf{90}, 062342 (2014)

\bibitem{Wang2016} Wang, C., Gao, Y.Y., Reinhold, P., Heeres, R.W., Ofek, N., Chou, K., Axline, C., Reagor, M., Blumoff, J., Sliwa, K.M., Frunzio, L., Girvin, S.M., Jiang, L., Mirrahimi, M., Devoret, M.H., Schoelkopf, R.J.: A Schr\"{o}dinger cat living in two boxes, Science 352, 1087 (2016)

\bibitem{James2000} James, D.F.V.: Quantum computation with hot and cold ions: an assessment of proposed schemes, Fortschr. Phys. \textbf{48}, 823 (2000)

\bibitem{DUAN} Duan, L.-M., Giedke, G., Cirac, J.I., Zoller, P.: Inseparability criterion for continuous variable systems, Phys. Rev. Lett. \textbf{84}, 2722 (2000)

\bibitem{KJG} Adesso, G., Ragy, S., Lee, A.R.: Continuous variable quantum information: Gaussian states and beyond, Open Syst. Inf. Dyn. \textbf{21}, 1440001 (2014)

\bibitem{Yoshihara2014} Yoshihara, F., Nakamura, Y., Yan, F., Gustavsson, S., Bylander, J., Oliver, W.D., Tsai, J.-S.: Flux qubit noise spectroscopy using Rabi oscillations under strong driving conditions, Phys. Rev. B \textbf{89}, 020503(R) (2014)

\bibitem{Pechal2014} Pechal, M., Huthmacher, L., Eichler, C., Zeytino\v{g}lu, S., Abdumalikov Jr.,  A.A., Berger, S., Wallraff, A., Filipp, S.: Microwave-controlled generation of shaped single photons in circuit quantum electrodynamics, Phys. Rev. X \textbf{4}, 041010 (2014)

\bibitem{Carmichael} Carmichael, H.J.: An open systems approach to quantum optics. Springer, Berlin (1993)

\bibitem{Beaudoin2011} Beaudoin, F., Gambetta, J.M., Blais, A.: Dissipation and ultrastrong coupling in circuit QED, Phys. Rev. A \textbf{84}, 043832 (2011)

\bibitem{Boi2009} Boissonneault, M., Gambetta, Blais, A.: Dispersive regime of circuit QED: Photon-dependent qubit dephasing and relaxation rates. Phys. Rev. A \textbf{79}, 013819 (2009)

\bibitem{JNATION} Johansson, J.R., Nation, P.D., Nori, F.: QuTiP: An open-source Python framework for the dynamics of open quantum systems, Comp. Phys. Comm. \textbf{183}, 1760 (2012); QuTiP 2: A Python framework for the dynamics of open quantum systems, \textit{ibid}. \textbf{184}, 1234 (2013)



\bibitem{Yan2016} Yan, F., Gustavsson, S., Kamal, A., Birenbaum, J., Sears, A.P., Hover, D., Gudmundsen, T.J., Rosenberg, D., Samach, G., Weber, S., Yoder, J.L., Orlando, T.P., Clarke, J., Kerman, A.J., Oliver, W.D.: The flux qubit revisited to enhance coherence and reproducibility, Nat. Commun. \textbf{7}, 12964 (2016)


\end{thebibliography}
\end{document}